\documentclass[12pt]{article}
\usepackage[margin=2cm]{geometry}
\usepackage{graphicx}
\usepackage{epsfig}
\usepackage{amsmath}
\usepackage{amssymb}
\usepackage{dcolumn}
\usepackage{bm}
\usepackage{amsmath}
\usepackage{enumitem}
\begin{titlepage}
\title{A Thousand Problems in Cosmology: Interaction in the Dark Sector}
\author{ 
Yu.L.~Bolotin ${}^{1}{}^*$ , V.A.~Cherkaskiy${}^{1}$,
O.A.~Lemets${}^{1}$, \\ I.V.~Tanatarov${}^{1,2}$, D.A.~Yerokhin${}^{1}$\\
{ }\\
\it ${}^1$A.I.Akhiezer Institute for Theoretical Physics,\\ \it
National Science Center "Kharkov Institute of Physics and
Technology",\\
\it Akademicheskaya Str. 1, 61108 Kharkov, Ukraine\\
\it ${}^{2}$V.N. Karazin Kharkov National University,\\
\it Svobody Sq. 4, 61077 Kharkov, Ukraine\\ {}\\ ${}^*$ybolotin@gmail.com
}
\date{}
\end{titlepage}
\begin{document}
\maketitle

The evolution of any broadly applied model is accompanied by multiple generalizations that aim to resolve conceptual difficulties and to explain the ever-growing pool of observational data. In the case of Standard Cosmological Model one of the most promising directions of generalization is replacement of the cosmological constant with a more complicated, dynamic, form of dark energy and incorporation of interaction between the dark components---dark energy (DE) and dark matter (DM). Typically, DE models are based on scalar fields minimally coupled to gravity, and do not implement explicit coupling of the field to the background DM. However, there is no fundamental reason for this assumption in the absence of an underlying symmetry which would suppress the coupling. Given that we do not know the true nature of either DE or DM, we cannot exclude the possibility that there is some kind of coupling between them. Whereas interactions between DE and normal matter particles are heavily constrained by observations (e.g. in the solar system and gravitational experiments on Earth), this is not the case for DM particles. In other words, it is possible for the dark components to interact with each other while not being coupled to standard model particles. Therefore, the possibility of DE-DM interaction should be investigated with utmost gravity.

\tableofcontents

\section{PHYSICAL MECHANISM OF ENERGY EXCHANGE}
\begin{enumerate}
\item \label{IDE_1}  Models with interaction between DM and the DE field can be realized if we make just an obvious assumption: the mass of the cold DM particles is a function of the DE field. Let the dark matter particles will be collisionless and nonrelativisic. Hence, the pressure of this fluid and its energy density are \(p_{dm}=0\) and \(\rho_{dm}=nm\) respectively, where $m$ is the rest mass and $n$ is the number density of DM particles. We define $m=\lambda\varphi$ where $\varphi$ is a scalar field and $\lambda$ is a dimensionless constant. Show how such assumption affects the scalar field dynamics \cite{0307350}.

\item \label{IDE_2}  Show that the DM on a inhomogeneous vacuum background can be treated as as interacting DE and DM \cite{1209.0563}.

\item \label{IDE_3}  Obtain general equations of motion for DE interacting with DM   \cite{1207.0250}.

\end{enumerate}

\section{PHENOMENOLOGY OF INTERACTING MODELS}
\begin{enumerate}[resume]
\item \label{IDE_4}  Find the effective state  parameters $w_{(de)eff}$ and $w_{(dm)eff}$ that would allow one to treat the interacting dark components as non-interacting.

\item \label{IDE_5}  Using the effective state parameters obtained in the previous problem, analyze dynamics of dark matter and dark energy depending on sign of the rate of energy density exchange in  the dark sector.

\item \label{IDE_6}Find the effective state  parameters $w_{(de)eff}$ and $w_{(dm)eff}$ for the case of the warm dark matter ($w_{dm}\ne0$) and analyze the features of dynamics in this case.

\item \label{IDE_7} Show that the quintessence coupled to DM with certain sign of the coupling constant behaves like a phantom uncoupled model, but without negative kinetic energy.

\item \label{IDE_8} In order to compare dynamics of a model with observational results it is useful to analyze all dynamic variables as functions of redshift rather than time. Obtain the corresponding transformation for the system of interacting dark components.

\item \label{IDE_9_0}Show that energy exchange between dark components leads to time-dependent effective potential energy term in the first Friedman equation. \cite{0502034}

\item \label{IDE_9}  Show that the system of interacting components can be treated as the uncoupled one due to introduction of the partial effective pressure of the dark components
     \[\Pi_{de}\equiv\frac{Q}{3H},\quad \Pi_{dm}\equiv-\frac{Q}{3H}.\]

\item \label{IDE_10}  Assume that the mass $m_{dm}$ of dark matter particles depends on a scalar field $\varphi$. Construct the model of interacting dark energy and dark  matter in this case.

\item \label{IDE_12n} Assume that the mass $m_{dm}$ of DM particles depends exponentially on the DE scalar field $m=m_*e^{-\lambda\varphi}$. Find the interaction term $Q$ in this case.

\item \label{IDE_11}  Find the equation of motion for the scalar field interacting with dark matter if its particles' mass depends on the scalar field.

\item \label{IDE_12}  Make the transformation from the variables $(\rho_{de}, \rho_{dm})$ to
\[\left(r=\frac{\rho_{dm}}{\rho_{de}}, \rho = \rho_{dm} + \rho_{de}\right)\] for the system of interacting dark components.
\item \label{IDE_13}  Generalize the result of previous problem to the case of warm dark matter.

\item \label{IDE_16n}  Calculate the derivatives $dr/dt$ and $dr/dH$ for the case of flat universe with the interaction $Q$.

\item \label{IDE_17n}  It was shown in the previous problem that
\[\dot r=r\left(\frac{\dot\rho_{dm}}{\rho_{dm}}-\frac{\dot\rho_{de}}{\rho_{de}}\right) = 3Hr \left(w_{de} +\frac{1+r}{\rho_{dm}}\frac Q{3H}\right)=(1+r)\left[3Hw_{de}\frac{r}{1+r}+\Gamma\right],\quad \Gamma\equiv\frac Q {\rho_{de}}.\] Exclude the interaction $Q$ and reformulate the equation in terms of $\rho_{de}$, $H$ and its derivatives.

\item \label{IDE_18n}

Generalize the result, obtained in the previous problem, for the case of non-flat Universe \cite{0606555}.

\item \label{IDE_14}  Show that critical points in the system of equations obtained in problem \ref{IDE_12} exist only for the case of dark energy of the phantom type.

\item \label{IDE_15}  Show that the result of previous problem holds also for warm dark matter.

\item \label{IDE_16}  Show that existence of critical points in the system of equations obtained in problem \ref{IDE_12} requires a transfer from dark energy to dark matter.

\item \label{IDE_17} Show that the result of previous problem holds also for warm dark matter.

\item \label{IDE_18}  Assume that the ratio of the interacting dark components equals \[r\equiv\frac{\rho_{dm}}{\rho_{de}}\propto a^{-\xi}, \quad \xi\ge0.\] Analyze how the interaction $Q$ depends on $\xi$.

\item \label{IDE_19}  Show that the choice
\[r\equiv\frac{\rho_{dm}}{\rho_{de}}\propto a^{-\xi}, \quad (\xi\ge0)\] guarantees existence of an early matter-dominated epoch.

\item \label{IDE_20}  Find the interaction $Q$ for the Universe with interacting dark energy and dark matter, assuming that ratio of their densities takes the form
    \[r\equiv\frac{\rho_{dm}}{\rho_{de}}=f(a),\] where $f(a)$ is an arbitrary differentiable function of the scale factor.

\item \label{IDE_26n}  Let \[Q=\frac{\dot f(t)}{f(t}\rho_{dm}.\] Show that the sign of the deceleration parameter is defined by the ratio \[\frac{\dot f}{fH}.\]

\item \label{IDE_27n}  Show that in the model, considered in the previous problem, the transition from the accelerated expansion to the decelerated one can occur only due to time dependence of the interaction.

\end{enumerate}
\section{SIMPLE LINEAR MODELS}
\begin{enumerate}[resume]
\item \label{IDE_21}  Find the scale factor dependence for the dark matter density assuming that the interaction between the dark matter and the dark energy equals $Q=\delta(a) H\rho_{dm}$.

\item \label{IDE_22}  Obtain the equation for the evolution of the DE  energy density for   $Q=\delta(a) H\rho_{dm}$.

\item \label{IDE_23}  Find $\rho_{dm}$ and $\rho_{de}$ in the case  $Q=\delta H\rho_{dm}$, $\delta=const$, $w_{de}=const$.

\item \label{IDE_24} As was shown above, interaction between dark matter and dark energy leads to non--conservation of matter, or equivalently, to scale dependence for the mass of particles that constitute the dark matter. Show that, within the framework of the model of previous problem ($Q=\delta H\rho_{dm}$, $\delta=const$, $w_{de}=const$) the relative change of particles mass per Hubble time equals to the interaction constant.

\item \label{IDE_25}  Find $\rho_{dm}$ and $\rho_{de}$ in the case  $Q=\delta H\rho_{de}$, $\delta=const$, $w_{de}=const$.

\item \label{IDE_26}  Find $\rho_{dm}$ and $\rho_{de}$ in the case  $Q=\delta(a) H\rho_{de}$, $\delta(a)=\beta_0a^\xi$, $w_{de}=const$ \cite{1111.6743}.

\item \label{IDE_27} Let's  look at a more general linear model for the expansion of a Universe that contains two interacting fluids with the equations of state
\[p_1 = (\gamma_1-1)\rho_1,\]	
\[p_2 = (\gamma_2-1)\rho_2,\]	
and energy exchange
	\[\dot\rho_1+3H\gamma_1\rho_1 = -\beta H\rho_1 + \alpha H\rho_2,\]
	\[\dot\rho_2+3H\gamma_2\rho_2 = \beta H\rho_1 - \alpha H\rho_2.\]
Here $\alpha$ and $\beta$ are constants describing the energy exchanges between the two fluids. Obtain the equation for $H(t)$ and find its solutions  \cite{9702029,0604063}).

\item \label{IDE_28}  Show that the energy balance equations (modified conservation equations) for $Q\propto H$ do not depend on H.

\item \label{IDE_36n} The Hubble parameter is present in the first Friedmann equation quadratically. This gives rise to a useful symmetry within a class of FLRW models. Because of this quadratic dependence, Friedmann's equation remains invariant under a transformation $H\to-H$ for the spatially flat case. This means it describes both expanding and contracting solutions. The transformation $H\to-H$ can be seen as a consequence of the change $a\to1/a$  of the scale factor of the FLRW metric. If, instead of just the first Friedmann equation, we want to make the whole system of Universe-describing equations invariant relative to this transformation, we must expand the set of values that undergo symmetry transformations. Then, when we refer to a duality transformation, we have in mind the following set of transformations
\[H\to\bar H=-H,\quad \rho\to\bar\rho=\rho,\quad p\to\bar p=-2\rho-p,\quad \gamma\equiv\frac{\rho+p}{\rho}\to\bar\gamma\equiv\frac{\bar\rho+\bar p}{\bar\rho}=-\gamma.\]

Generalize the duality transformation to the case of interacting components \cite{0505096}.

\end{enumerate}

\section{COSMOLOGICAL MODELS WITH A CHANGE OF THE DIRECTION OF ENERGY TRANSFER}

{\it Let us consider one more type of interaction $Q$, whose sign (i.e., the direction of energy transfer) changes when the mode of decelerated expansion is replaced by the mode of accelerated expansion, and vice versa. The simplest interaction of this type is the one proportional to the deceleration parameter. An example of such interaction is
\[Q=q(\alpha\dot\rho+\beta H\rho),\]
where $\alpha$ and $\beta$ are dimensionless constants, and $\rho$ can be any of densities $\rho_{de}$, $\rho_{dm}$ or $\rho_{tot}$. In the following problems for simplicity we restrict ourselves to the decaying $\Lambda$ model, for which $\dot\rho_{de}=\dot\rho_\Lambda=-Q$ and $p_{de}=-\rho_{de}$. (after  \cite{1010.1074})}
\begin{enumerate}[resume]
\item \label{IDE_29} Construct general procedure to the Hubble parameter and the deceleration parameter for the case \(Q=q(\alpha\dot\rho_{dm}+\beta H\rho_{dm})\).

\item \label{IDE_30}  Find deceleration parameter for the case $\alpha=0$ in the model considered in the previous problem.

\item \label{IDE_31}  Consider the model $Q=q(\alpha\dot\rho_\Lambda+3\beta H\rho_\Lambda)$. Obtain the Hubble parameter $H(a)$ and deceleration parameter for the case $\alpha=0$.
\end{enumerate}

\section{NON-LINEAR INTERACTION IN THE DARK SECTOR}
{\it The interaction studied so far are linear in the sense that the interaction term in the individual energy balance equations is proportional either to dark matter density or to dark energy density or to a linear combination of both densities. Also from a physical point of view an interaction proportional to the product of dark components seems preferred: an interaction between two components should depend on the product of the abundances of the individual components, as, e.g., in chemical reactions. Moreover, such type of interaction looks more preferable then the linear one when compared with the observations. Below we investigate the dynamics for a simple two-component model with a number of non-linear interactions.}

In problems \ref{IDE_32}-\ref{IDE_35} let a spatially flat FLRW Universe contain perfect fluids with densities $\rho_1$ and $\rho_2$. Consider a nonlinear interaction of the form $Q=\gamma\rho_1\rho_2$ \cite{1009.4942}

\begin{enumerate}[resume]
\item \label{IDE_32}  Consider a model in which both fluids are dust. Find $r(t)\equiv\rho_1(t)/\rho_2(t)$.

\item \label{IDE_33}  Consider a Universe with more than two CDM components interacting with each other. What is the asymptotic behavior of the individual densities of the components in the limit $t\to\infty$?

\item \label{IDE_34} Consider a Universe containing a cold dark matter  and a dark energy, in which the dark energy behaves like a cosmological constant.  Show that in such model dark energy is a perpetual component of the Universe.

\item \label{IDE_35} Consider a two-component Universe with the interaction $Q=\gamma\rho_1\rho_2$. Let one component is CDM ($\rho_1=\rho_{dm}$, $w_1=0$), and the second is the dark energy with arbitrary state equation ($\rho_2=\rho_{de}$, $w_2=\gamma_{de}-1$). (The case considered in the previous problem corresponds to $\gamma_{de}=0$.) Find the relation between the dark energy and dark matter densities.

\item \label{IDE_36} Let interaction term $Q$ be a non-linear function of the energy densities of the components and/or the total energy density. Motivated by the structure
\[\rho_{dm} = \frac r{1+r}\rho, \quad \rho_{de} = \frac r{1+r}\rho,\]
\[\rho\equiv\rho_{dm}+\rho_{de},\quad r\equiv\frac{\rho_{dm}}{\rho_{de}}\]
consider ansatz
	\[Q=3H\gamma\rho^m r^n(1+r)^s.\]

where $\gamma$ is a positive coupling constant. Show that for $s=-m$ interaction term is proportional to a power of products of the densities of the components. For $(m,n,s)=(1,1,-1)$ and $(m,n,s)=(1,0,-1)$ reproduce the linear case  \cite{1112.5095}.

\item \label{IDE_37} Find analytical solution of non-linear interaction model covered by the ansatz of previous problem for $(m,n,s)=(1,1,-2)$, $Q=3H\gamma\rho_{de}\rho_{dm}/\rho$

\item \label{IDE_38} Find analytical solution of non-linear interaction model for $(m,n,s)=(1,2,-2)$, $Q=3H\gamma\rho_{dm}^2/\rho$.

\item \label{IDE_39}  Find analytical solution of non-linear interaction model for $(m,n,s)=(1,0,-2)$, $Q=3H\gamma\rho_{de}^2/\rho$.

\item \label{IDE_40}  Consider a flat Universe filled by CDM and DE with a polytropic equation of state
	\[p_{de}=K\rho_{de}^{1\frac1n}\]
where $K$ and $n$ are the polytropic constant and polytropic index, respectively. Find dependence of DE density on the scale factor under assumption that the interaction between the dark components is $Q=3\alpha H\rho_{de}$ \cite{1012.2692}.

\item \label{IDE_41} Show that under certain conditions the interacting polytropic dark energy with $Q=3\alpha H\rho_{de}$ behaves as the phantom energy.

\item \label{IDE_42}  Find deceleration parameter for the system considered in the problem \ref{IDE_40}

\end{enumerate}

\section{PHASE SPACE STRUCTURE OF MODELS WITH INTERACTION}
{\it The evolution of a Universe filled with interacting components can be effectively analyzed in terms of dynamical systems theory. Let us consider the following coupled differential equations for two variables
\begin{equation}
\label{IDE_s6_1}
\begin{array}{l}
\dot x=f(x,y,t),\\
\dot y=g(x,y,t).
\end{array}
\end{equation}	
We will be interested in the so-called autonomous systems, for which the functions $f$ and $g$ do not contain explicit time-dependent terms.
A point $(x_c,y_c)$ is said to be a fixed (a.k.a. critical) point of the autonomous system if
	\[f(x_c,y_c)=g(x_c,y_c)=0.\]
A critical point $(x_c,y_c)$ is called an attractor when it satisfies the condition \(\left(x(t),y(t)\right)\to(x_c,y_c)\) for $t\to\infty$.
Let's look at the behavior of the dynamical system (\ref{IDE_s6_1}) near the critical point. For this purpose, let us consider small perturbations around the critical point
	\[x=x_c+\delta x,\quad y=y_c+\delta y.\]
Substituting it into (\ref{IDE_s6_1}) leads to the first-order differential equations:
	\[\frac{d}{dN}\left(\begin{array}{c}\delta x\\ \delta y\end{array}\right) = \hat M \left(\begin{array}{c}\delta x\\ \delta y\end{array}\right).\]
Taking into account the specifics of the problem that we are solving, we made the change \[\frac{d}{dt}\to\frac{d}{dN},\]
where $N=\ln a$. The  matrix $\hat M$ is given by
\[\hat M =
\left(
\begin{array}{lr}
\frac{\partial f}{\partial x} & \frac{\partial f}{\partial y} \\
{} & {}\\
\frac{\partial g}{\partial x} & \frac{\partial g}{\partial y}
\end{array}
\right)
\]
The general solution for the linear perturbations reads
	\[\delta x=C_1e^{\lambda_1 N} + C_2e^{\lambda_2 N},\]
	\[\delta y=C_3e^{\lambda_1 N} + C_4e^{\lambda_2 N},\]
The stability around the fixed points depends on the nature of the eigenvalues.

Let us treat the interacting dark components as a dynamical system described by the equations
\[\rho'_{de}+3(1+w_{de})\rho_{de}=-Q\]
\[\rho'_{dm}+3(1+w_{dm})\rho_{dm}=Q\]
Here, the prime denotes the derivative with respect to $N=\ln a$. Note that although the interaction can significantly change the cosmological evolution, the system is still autonomous. We consider the following specific interaction forms, which were already analyzed above:
\[Q_1=3\gamma_{dm}\rho_{dm},\quad Q_1=3\gamma_{de}\rho_{de},\quad Q_1=3\gamma_{tot}\rho_{tot}\]}
\begin{enumerate}[resume]

\item \label{IDE_43}  Find effective EoS parameters $w_{(dm)eff}$ and $w_{(de)eff}$ for the interactions $Q_1$, $Q_2$ and $Q_3$.
\item \label{IDE_52n} Find the critical points of equation for ratio $r=\rho_{dm}/\rho_{de}$ if $Q=3\alpha H(\rho_{dm}+\rho_{de})$, where the phenomenological parameter $\alpha$ is a dimensionless, positive constant, $w_{dm}=0$, $w_{de}=const$.

\item \label{IDE_53n}  Show, that the remarkable property of the model, considered in the previous problem,
is that for the interaction parameter $\alpha$, consistent with the current observations $\alpha<2.3\times10^{-3}$ the ratio $r$ tends to a stationary but unstable value at early times, $r_s^+$, and to a stationary and stable value, $r_s^-$ (an attractor), at late times. Consequently, as the Universe expands, $r(a)$ smoothly evolves from $r_s^+$ to the attractor solution $r_s^-$.

\item \label{IDE_44} Transform the system of equations
	\[\rho'_{de}+3(1+w_{de})\rho_{de}=-Q,\]
	\[\rho'_{dm}+3(1+w_{dm})\rho_{dm}=Q,\]
into the one for the fractional density energies \cite{1003.2788}.

\item \label{IDE_45} Analyze the critical points of the autonomous system, obtained in the previous problem

\item \label{IDE_46}  Construct the stability matrix for the dynamical system considered in the problem \ref{IDE_44} and determine its eigenvalues.

\item \label{IDE_47}  Using result of the previous problem, determine eigenvalues of the stability matrix for the following cases: i) $\Omega_{dm} = 1$, $\Omega_{de} = 0$, $f_j \ne 0$; ii) $\Omega_{dm} = 0$, $\Omega_{de} = 1$, $f_j \ne 0$; iii) $f_j = 0$.

\item \label{IDE_48}  Obtain position and type of the critical points obtained in the previous problem for the case of cosmological constant interacting with dark matter as $Q=3\gamma_{dm}\rho_{dm}$.

\item \label{IDE_49} Construct the stability matrix for the following dynamical system
\begin{align}
\nonumber \rho' & = - \left(1+\frac{w_{de}}{1+r}\right)\rho,\\
\nonumber r' & = r \left[w_{de} - \frac{(1+r)^2}{r\rho}\Pi\right],
\end{align}
and determine its eigenvalues\cite{1112.5095}.

\end{enumerate}

\section{PECULIARITY OF  DYNAMICS OF SCALAR FIELD COUPLED TO DARK  MATTER}
\subsection{Interacting quintessence model}
{\it Given that the quintessence field and the dark matter have unknown physical natures, there seem to be no a priori reasons to exclude a coupling between the two components. Let us consider a two-component system (scalar field $\varphi$ + dark matter) with the energy density and pressure
\[\rho=\rho_\varphi+\rho_{dm},\quad p=p_\varphi+p_{dm}\]
(we do not exclude the possibility of warm DM ($p_{dm}\ne0$).)
If some interaction exists between the scalar field and DM, then
\[\dot\rho_{dm}+3H(\rho_{dm}+p_{dm})=Q\]
\[\dot\rho_\varphi+3H(\rho_\varphi+p_\varphi)=-Q.\]
Using the effective pressures $\Pi_\varphi$ and $\Pi_{dm}$,
\[Q=-3H\Pi_{dm}=3H\Pi_\varphi\]
one can transit to the system
\begin{align}
\nonumber
\dot\rho_{dm}+3H(\rho_{dm}+p_{dm}+\Pi_{dm}) & =0,\\
\nonumber
\dot\rho_\varphi+3H(\rho_\varphi+p_\varphi+\Pi_\varphi) & =0.
\end{align}}
\begin{enumerate}[resume]
\item \label{IDE_60}  Obtain the modified Klein-Gordon equation for the scalar field interacting with the dark matter.

\item \label{IDE_61}     Consider a quintessence scalar field $\varphi$ which couples to the dark matter via, e.g., a Yukawa-like interaction $f(\varphi/M_{Pl})\bar\psi\psi$, where $f$ is an arbitrary function of $\varphi$ and $\psi$ is a dark matter Dirac spinor. Obtain the modified Klein-Gordon equation for the scalar field interacting with the dark matter in such way \cite{0510628v2}.

\item \label{IDE_62} The problems \ref{IDE_62}-\ref{IDE_66} are inspired by \cite{0105479}.

    Show that the Friedman equation with interacting scalar field and dark matter allow existence of stationary solution for the ratio $r\equiv\rho_{dm}/\rho_\varphi$.

\item \label{IDE_63}  Find the form of interaction $Q$ which provides the stationary relation $r$ for interacting cold dark matter and quintessence in spatially flat Universe.

\item \label{IDE_64} For the interaction $Q$ which provides the stationary relation $r$ for interacting cold dark matter and quintessence in spatially flat Universe (see the previous problem), find the dependence of $\rho_{dm}$ and $\rho_\varphi$ on the scale factor.

\item \label{IDE_65} Show that in the case of interaction $Q$ obtained in the problem \ref{IDE_63}, the scalar field $\varphi$ evolves logarithmically with time.

\item \label{IDE_66}  Reconstruct the potential $V(\varphi)$, which realizes the solution $r=const$, obtained in the problem \ref{IDE_63}.

\item \label{IDE_67}Let the DM particle's mass $M$ depend exponentially on the DE scalar field as $M=M_*e^{-\lambda\varphi}$, where $\lambda$ is positive constant and the scalar field potential is
\[V(\varphi)=V_* e^{\eta\varphi}.\]
Obtain the modified Klein-Gordon equation for this case\cite{0503075}.

\item \label{IDE_68} Let the DM particle's mass $M$ depend exponentially on the DE scalar field
as $M=M_*e^{-\alpha}$, and the scalar field potential is
\[V(\varphi)=V_* e^{\beta},\]
where $\alpha,\beta>0$. Obtain the modified Klein-Gordon equation for this case.

\end{enumerate}
\subsection{Interacting Phantom}
{\it Let the Universe contain only noninteracting cold dark matter ($w_{dm}=0$) and a phantom field ($w_{de}<-1$). The densities of these components evolve separately: $\rho_{dm}\propto a^{-3}$ and $\rho_{de}\propto a^{-3(1+w_{de})}$.  If matter domination ends at $t_m$, then at the moment of time \[t_{BR}=\frac{w_{de}}{1+w_{de}}t_m\] the scale factor, as well as a series of other cosmological characteristics of the Universe become infinite. This catastrophe has earned the name "Big Rip". One of the way to avoid the unwanted big rip singularity is to allow for a suitable interaction between the phantom energy and the background dark matter.}
\begin{enumerate}[resume]
\item \label{IDE_69} Show that through a special choice of interaction, one can mitigate the rise of the phantom component and make it so that components decrease with time if there is a transfer of energy from the phantom field to the dark matter. Consider case of $Q=\delta(a)H\rho_{dm}$ and $w_{de}=const$.

\item \label{IDE_70} Calculate the deceleration parameter for the model considered in the previous problem.

\item \label{IDE_71}Let the interaction $Q$ of phantom field $\varphi$ with DM provide constant relation $r=\rho_{dm}/\rho_\varphi$. Assuming that $w_\varphi=const$, find $\rho_\varphi(a)$, $\rho_\varphi(\varphi)$ and $a(\varphi)$ for the case of cold dark matter (CDM)\cite{0411524}.

\item \label{IDE_72}  Construct the scalar field potential, which realizes the given relation $r$ for the model considered in the previous problem.

\end{enumerate}
\subsection{Tachyonic Interacting Scalar Field}
{\it Let us consider a flat Friedmann Universe filled with a spatially homogeneous tachyon field $T$ evolving according to the Lagrangian
\[L=-V(T)\sqrt{1-g_{00}\dot T^2}.\]
The energy density and the pressure of this field are, respectively
\[\rho_T=\frac{V(T)}{\sqrt{1-\dot T^2}}\]
and
\[p_T=-V(T)\sqrt{1-\dot T^2}.\]
The equation of motion for the tachyon is
\[\frac{\ddot T}{1-\dot T^2}+3H\dot T+\frac{1}{V(T)}\frac{dV}{dT}.\]
(Problems \ref{IDE_73}-\ref{IDE_77} are after  \cite{0404086}.)}
\begin{enumerate}[resume]
\item \label{IDE_73}  Find interaction of tachyon field with cold dark matter (CDM), which results in $r\equiv\rho_{dm}/\rho_T=const$.

\item \label{IDE_74}   Show that the stationary solution $\dot r=0$ exists only when the energy of the tachyon field  is transferred to the dark matter.

\item \label{IDE_75} Find the modified Klein-Gordon equation for arbitrary interaction $Q$ of tachyon scalar field with dark matter.

\item \label{IDE_76}  Find the modified Klein-Gordon equation for the interaction $Q$ obtained in the problem \ref{IDE_73} and obtain its solutions for the case $\dot\varphi=const$.

\item \label{IDE_77}  Show that sufficiently small values of tachyon field provide the accelerated expansion of Universe.

\end{enumerate}
\section{REALIZATION OF INTERACTION IN THE DARK SECTOR}
\subsection{Vacuum Decay into Cold Dark Matter}
{\it Let us consider the Einstein field equations
\[R_{\mu\nu}-frac12Rg_{\mu\nu}=8\pi G\left(T_{\mu\nu}+\frac\Lambda{8\pi G}g_{\mu\nu}\right).\]	
According to the Bianchi identities, (i) vacuum decay is possible only from a previous existence of some sort of non-vanishing matter and/or radiation, and (ii) the presence of a time-varying cosmological term results in a coupling between $T_{\mu\nu}$ and $\Lambda$. We will assume (unless stated otherwise) coupling only between vacuum and CDM particles, so that
	\[u_\mu,T_{;\nu}^{(CDM)\mu\nu}=-u_\mu\left(\frac{\Lambda g^{\mu\nu}}{8\pi G}\right)_{;\nu}= -u_\mu\left(\rho_\Lambda g^{\mu\nu}\right)_{;\nu}\]
where $T_{\mu\nu}^{(CDM)}=\rho_{dm}u^\mu u^\nu$ is the energy-momentum tensor of the CDM matter and $\rho_\Lambda$ is the vacuum energy density. It immediately follows that
	\[\dot\rho_{dm}+3H\rho_{dm}=-\dot\rho_\Lambda.\]
Note that although the vacuum is decaying, $w_\Lambda=-1$ is still constant, the physical equation of state (EoS) of the vacuum $w_\Lambda\equiv=p_\Lambda/\rho_\Lambda$ is still equal to constant $-1$, which follows from the definition  of the cosmological constant.

(see  \cite{0408495,0507372} )}
\begin{enumerate}[resume]
\item \label{IDE_78} Since vacuum energy is constantly decaying into CDM, CDM will dilute in a smaller rate compared with the standard relation $\rho_{dm}\propto a^{-3}$. Thus we assume that $\rho_{dm}=\rho_{dm0}a^{-3+\varepsilon}$, where $\varepsilon$ is a small positive small constant. Find the dependence $\rho_\Lambda(a)$ in this model.

\item \label{IDE_79} Solve the previous problem for the case when vacuum energy is constantly decaying into radiation.

\item \label{IDE_80} Show that existence of a radiation dominated stage is always guaranteed in scenarios, considered in the previous problem.

\item \label{IDE_81}  Find how the new temperature law scales with redshift in the case of vacuum energy decaying into radiation.

\end{enumerate}
{\it Since the energy density of the cold dark matter is $\rho_{dm}=nm$, there are two possibilities for storage of the energy received from the vacuum decay process:

(i) the equation describing concentration, $n$, has a source term while the proper mass of CDM particles remains constant;

(ii) the mass $m$ of the CDM particles is itself a time-dependent quantity, while the total number of CDM particles, $N=na^3$, remains constant.

Let us consider both the possibilities.}
\subsection{Vacuum decay into CDM particles}
\begin{enumerate}[resume]
\item \label{IDE_82} Find dependence of total particle number on the scale factor in the model considered in problem \ref{IDE_78}.

\item \label{IDE_83} Find time dependence of CDM particle mass in the case when there is no creation of CDM particles in the model considered in problem \ref{IDE_78}.

\item \label{IDE_84} Consider a model where the cosmological constant $\Lambda$ depends on time as $\Lambda=\sigma H$. Let a flat Universe be filled by the time-dependent cosmological constant and a component with the state equation $p_\gamma= (\gamma-1)\rho_\gamma$. Find solutions of Friedman equations for this system \cite{0711.2686} .

\item \label{IDE_85}  Show that the model considered in the previous problem correctly reproduces the scale factor evolution both in the radiation-dominated and non-relativistic matter (dust) dominated cases.

\item \label{IDE_86}  Find the dependencies $\rho_\gamma(a)$ and $\Lambda(a)$ both in the radiation-dominated and non-relativistic matter dominated cases in the model considered in problem \ref{IDE_84}.

\item \label{IDE_87} Show that for the $\Lambda(t)$ models \[T\frac{dS}{dt}=-\dot\rho_\Lambda a^3.\]

\item \label{IDE_87_2}  Consider a two-component Universe filled by matter with the state equation $p=w\rho$ and cosmological constant and rewrite the second Friedman equation in the following form

\item \label{IDE_87_3}  Consider a two-component Universe filled by matter with the state equation $p=w\rho$ and cosmological constant with quadratic time dependence $\Lambda(\tau)=\mathcal{A}\tau^2$ and find the time dependence for of the scale factor.

\item \label{tauell}  Consider a flat two-component Universe filled by matter with the state equation $p=w\rho$ and cosmological constant with quadratic time dependence $\Lambda(\tau)=\mathcal{A}\tau^{\ell}$. Obtain the differential equation for Hubble parameter in this model and classify it.

\item \label{IDE_87_5} Find solution of the equation obtained in the previous problem in the case ${\ell =1}$. Analyze the obtained solution.

\item \label{IDE_87_6}  Solve the equation obtained in the problem \ref{tauell} for ${\ell =2}.$ Consider the following cases
\begin{description}
 \item[a)] $\lambda_0 > -1/(3\gamma\tau_0)^2,$
 \item[b)] $\lambda_0 = -1/(3\gamma\tau_0)^2,$
 \item[c)] $\lambda_0 < -1/(3\gamma\tau_0)^2$
\end{description}
(see the previous problem). Analyze the obtained solution.

\item \label{IDE_87_7}  Consider a flat two-component Universe filled by matter with the state equation $p=w\rho$ and cosmological constant with the following scale factor dependence
\begin{equation}
\Lambda = {\cal B} \, a^{-m}.
\label{Bam}
\end{equation}
Find dependence of energy density of matter on the scale factor in this model.

\item \label{IDE_87_8}  Find dependence of deceleration parameter on the scale factor for the model of previous problem.

\end{enumerate}
\section{STATEFINDER PARAMETERS FOR INTERACTING DARK ENERGY AND COLD DARK MATTER.}
\begin{enumerate}[resume]
\item \label{IDE_88}

    Show that in flat Universe both the Hubble parameter and deceleration parameter do not depend on whether or not dark components are interacting. Become convinced the second derivative $\ddot H$ does depend on the interaction between the components\cite{0311067}.

\item \label{IDE_89} Find  statefinder parameters for interacting dark energy and cold dark matter.

\item \label{IDE_90} Show that the statefinder parameter $r$ is generally necessary to characterize any variation in the overall equation of state of the cosmic medium.

\item \label{IDE_91}  Find relation between the statefinder parameters in the flat Universe.

\item \label{IDE_92} Express the statefinder parameters in terms of effective state parameter $w_{(de)eff}$, for which \[\dot\rho_{de}+3H(1+w_{(de)eff})\rho_{de}=0.\]

\item \label{IDE_93} Find the statefinder parameters for $Q=3\delta H\rho_{dm}$, assuming that $w_{de}=const$.

\item \label{IDE_94} Find statefinder parameters for the case $\rho_{dm}/\rho_{de}=a^{-\xi}$, where $\xi$ is a constant parameter in the range $[0,3]$ and $w_{de}=const$.

\item \label{IDE_95} Show that in the case $\rho_{dm}/\rho_{de}=a^{-\xi}$ the current value of the statefinder parameter $s=s_0$ can be used to measure the deviation of cosmological models from the SCM.

\item \label{IDE_96} Find how the statefinder parameters enter the expression for the luminosity distance.

\end{enumerate}
\section{INTERACTING HOLOGRAPHIC DARK ENERGY}
{\it The traditional point of view assumed that dominating part of
degrees of freedom in our World are attributed to physical fields.
However it became clear soon that such concept complicates
the construction of Quantum Gravity: it is necessary to introduce small
distance cutoffs for all integrals in the theory in order to make it
sensible. As a consequence, our World should be described on a
three-dimensional discrete lattice with the period of the order of Planck
length. Lately some physicists share an even more radical
point of view: instead of the three-dimensional lattice, complete
description of Nature requires only a two-dimensional one, situated on
the space boundary of our World. This approach is based on the
so-called ''holographic principle''. The name is related to the optical
hologram, which is essentially a two-dimensional record of a
three-dimensional object. The holographic principle consists of two
main statements:
\begin{enumerate}
\item All information contained in some region of space can be
''recorded'' (represented) on the boundary of that region.
\item The theory, formulated on the boundaries of the considered
region of space, must have no more than one degree of freedom per
Planck area:
\begin{equation}
\label{Hol_f:1}
    N\le \frac{A}{A_{pl}},\quad A_{pl}=\frac{G\hbar}{c^3}.
\end{equation}
\end{enumerate}
Thus, the key piece in the holographic principle is
the assumption that all the information about the Universe can be
encoded on some two-dimensional surface --- the holographic
screen. Such approach leads to a new interpretation of cosmological
acceleration and to an absolutely unusual understanding of Gravity. The
Gravity is understood as an entropy force, caused by variation of
information connected to positions of material bodies. More
precisely, the quantity of information related to matter and its
position is measured in terms of entropy. Relation between the
entropy and the information states that the information change is
exactly the negative entropy change $\Delta I=-\Delta S$. Entropy change
due to matter displacement leads to the so-called entropy force, which,
as will be proven below, has the form of gravity. Its origin
therefore lies in the universal tendency of any macroscopic theory
to maximize the entropy. The dynamics can be constructed in terms
of entropy variation and it does not depend on the details of
microscopic theory. In particular, there is no fundamental field
associated with the entropy force. The entropy forces are typical
for macroscopic systems like colloids and biophysical systems. Big
colloid molecules, placed in thermal environment of smaller
particles, feel the entropy forces. Osmose is another phenomenon
governed by the entropy forces.

Probably the best known example of the entropy force is the elasticity of a polymer
molecule. A single polymer molecule can be modeled as a
composition of many monomers of fixed length. Each monomer can
freely rotate around the fixation point and choose any spacial
direction. Each of such configurations has the same energy. When the
polymer molecule is placed into a thermal bath, it prefers to form a
ring as the entropically most preferable configuration: there are
many more such configurations when the polymer molecule is short,
than those when it is stretched. The statistical tendency to transit
into the maximum entropy state transforms into the macroscopic
force, in the considered case---into the elastic force.

Let us consider a small piece of holographic screen and a particle
of mass $m$ approaching it. According to the holographic principle,
the particle affects the amount of the information (and therefore of the
entropy) stored on the screen. It is natural to assume that
entropy variation near the screen is linear on the displacement
$\Delta x$:
\begin{equation}
\Delta S = 2\pi k_B \frac{mc}{\hbar} \Delta x. \label{delta_s}
\end{equation}
The factor $2\pi$ is introduced for convenience, which the reader
will appreciate solving the problems of this section. In order to
understand why this quantity should be proportional to mass,
let us imagine that the particle has split into two or more particles
of smaller mass. Each of those particles produces its own entropy
change when displaced by $\Delta x$. As entropy and mass are both
additive, then it is natural that the former is proportional to the
latter. According to the first law of thermodynamics, the entropy
force related to information variation satisfies the equation
\begin{equation}
F\Delta x = T\Delta S. \label{delta_x}
\end{equation}
If we know the entropy gradient, which can be found from (\ref{delta_s}),
and the screen temperature, we can calculate the entropy
force.

An observer moving with acceleration $a$, feels the
temperature (the Unruh temperature)
\begin{equation}
\label{Hol_f_Unruh:4}
k_B T_U=\frac{1}{2\pi}\frac\hbar c a.
\end{equation}
Let us assume that the total energy of the system equals $E$. Let us
make a simple assumption that the energy is uniformly distributed
over all $N$ bits of information on the holographic screen. The
temperature is then defined as the average energy per bit:
\begin{equation}
E =\frac12 N k_B T. \label{average_e}
\end{equation}
Equations (\ref{delta_s})--(\ref{average_e}) allow one to describe the holographic
dynamics, and as a particular case---the dynamics of the Universe, and
all that without the notion of Gravity.}
\begin{enumerate}[resume]

\item\label{IDE_97_1} For the interacting holographic dark energy $Q = 3\alpha H \rho_L,$ with the Hubble radius as the IR cutoff, find the depending on the time for the scale factor, the Hubble parameter and the deceleration parameter.

\item \label{IDE_97} Show that for the choice $\rho_{hde}\propto H^2$ ($\rho_{hde}=\beta H^2$, $\beta=const$)an interaction is the only way to have an equation of state different from that of the dust.

\item \label{IDE_98}    Calculate the derivative \[\frac{d\rho_{de}}{d\ln a}\] for the holographic dark energy model, where IR cut-off $L$ is chosen to be equal to the future event horizon\cite{0711.1641}.

\item \label{IDE_99} Find the effective state parameter value $w_{eff}$, such that \[\rho'_{de}+3(1+w_{eff})\rho_{de}=0\] for the holographic dark energy model, considered in the previous problem, with the interaction of the form $Q=3\alpha H\rho_{de}$

\item \label{IDE_100} Analyze how fate of the Universe depends on the parameter $c$ in the holographic dark energy model, where IR cut-off $L$ is chosen to be equal to the future event horizon.

\item \label{IDE_101} In the case of interacting holographic Ricci dark energy with interaction is given by
 \begin{eqnarray}
 \label{intrate}
   Q=\gamma H \rho_{_{\cal R}},
  \end{eqnarray}
  where $\gamma$ is a dimensionless parameter,
find the dependence of the density of dark energy and dark matter on the scale factor.

\item \label{IDE_102}  Find the exact solutions for linear interactions between Ricci DE and DM, if the energy density of Ricci DE is given by $ \rho_x =\left(2\dot H + 3\alpha H^2\right)/\Delta,$  where $\Delta=\alpha -\beta$ and $\alpha,\,\beta$ are constants.

\item\label{IDE_103} Find  the equation of motion for the relative density

  \end{enumerate}
\section{TRANSIENT ACCELERATION}

{\it Unlike fundamental theories, physical models only reflect the current state of our understanding of a process or phenomenon for the description of which they were developed. The efficiency of a model is to a significant extent determined by its flexibility, i.e., its ability to update when new information appears. Precisely for this reason, the evolution of any broadly applied model is accompanied by numerous generalizations aimed at resolving conceptual problems, as well as a description of the ever increasing number of observations. In the case of the SCM, these generalizations can be divided into two main classes. The first is composed of generalizations that replace the cosmological constant with more complicated dynamic forms of DE, for which the possibility of their interaction with DM must be taken into account. Generalizations pertaining to the second class are of a more radical character. The ultimate goal of these generalizations (explicit or latent) consists in the complete renunciation of dark components by means of modifying Einstein's equations. The generalizations of both the first and second classes can be demonstrated by means of a phenomenon that has been termed ``transient acceleration''.

A characteristic feature of the dependency of the deceleration parameter $q$ on the redshift $z$ in the SCM is that it monotonically tends to its limit value $q(z) = —1$ as $z \to — 1.$ Physically, this means that when DE became the dominant component (at $z \sim 1$), the Universe in the SCM was doomed to experience eternal accelerating expansion.
In what follows, we consider several cosmological models that involve dynamic forms of DE that lead to transition acceleration, and we also discuss what the observational data says about the modern rate of expansion of the Universe.

Barrow \cite{Barrow} was among the first to indicate that transient acceleration is possible in principle. He showed that within quite sound scenarios that explain the current accelerated expansion of the Universe, the possibility was not excluded of a return to the era of domination of nonrelativistic matter and, consequently, to decelerating expansion. Therefore, the transition to accelerating expansion does not necessarily mean eternal accelerating expansion. Moreover, in Barrow's article, it was shown to be neither the only possible nor the most probable course of events.}
\begin{enumerate}[resume]
\item  Consider a simple model of transient acceleration with decaying cosmological constant
\begin{equation} \label{ec}
\dot{\rho}_{m} + 3\frac{\dot{a}}{a}\rho_{m} = - \dot{\rho}_\Lambda\;,
\end{equation}
where $\rho_{m}$ and $\rho_\Lambda$ energy density DE and cosmological constant  $\Lambda$.

At the early stages of the expansion of the Universe, when $\rho_\Lambda$ is quite small, such a decay does not influence cosmological evolution in any way. At later stages, as the DE contribution increases, its decay has an ever increasing effect on the standard dependence of the DM energy density $\rho_{m} \propto a^{-3}$ on the scale factor $a$. We consider the deviation to be described by a function  of the scale factor - $\epsilon(a)$.
\begin{equation} \label{dm}
\rho_{m} = \rho_{m, 0}a^{-3 + \epsilon(a)}\;,
\end{equation}
where $a_0 = 1$ in the present epoch.
Other fields of matter (radiation, baryons) evolve independently and are conserved. Hence, the DE density has the form
\begin{equation}\label{decayv}
\rho_{\Lambda} =  \rho_{m0} \int\limits_{a}^{1}\frac{\epsilon(\tilde{a}) + \tilde{a}\epsilon' \ln(\tilde{a})}{\tilde{a}^{4 - \epsilon(\tilde a)}} d\tilde{a} + {\rm{X}}\;,
\end{equation}
where the prime denotes the derivative with respect to the scale factor, and ${\rm{X}}$ is the integration constant. If radiation is neglected, the first Friedmann equation takes the form
\begin{equation}
\label{friedmann} {{H}}= H_0\left[\Omega_{b,0}{a}^{-3} + \Omega_{m0}\varphi(a) + {\Omega}_{{\rm{X,0}}}\right]^{1/2},
\end{equation}
Using the assumption that the function $\epsilon (a)$ has the following simple form
\begin{equation}
\label{Parametrization_a}
\epsilon(a) = \epsilon_0a^\xi\ = \epsilon_0(1+z)^{-\xi},
\end{equation}
where $\epsilon_0$ and $\xi$ can take both positive and negative values, find function $\varphi(a)$ and relative energy density $\Omega_b(a)$, $\Omega_{m}(a)$ and $\Omega_{\Lambda}(a)$.

  \item Using the results of the previous problem, find the deceleration parameter for this model is $q(a)$. Draw the graph deceleration parameter as a function of $\log(a)$ for various values of $\epsilon_0$ and $\xi$: $\xi = 1.0$ and $\epsilon_0 = 0.1$, $\xi = -1.0$ and $\epsilon_0 = 0.1$, $\xi = 0.8$ and $\epsilon_0 = 0.5$, $\xi = -0.5$ and $\epsilon_0 = -0.1.$.

\item Consider the possibility of an accelerating transient regime within the interacting scalar field model using potential of the form
\begin{equation}
\label{V_fransient}
V(\phi)=\rho_{\phi\,0}[1-\frac{\lambda}{6}(1+\alpha \sqrt{\sigma}\phi)^2)]
\exp{[-\lambda\sqrt{\sigma}(\phi+ \frac{\alpha \sqrt{\sigma}}{2}\phi^2)]},
\end{equation}
where $\rho_{\phi\,0}$ is a constant energy density, $\sigma=8\pi G/\lambda$, and $\alpha$ and $\lambda$ are two dimensionless, positive parameters of the model, that the deceleration parameter is non-monotonically dependent on the scale factor.
Plot the deceleration parameter as a function of the scale factor.

\item Consider a simple parameterization:
\begin{equation}
\label{Qsimple}
Q=3\beta(a)H \rho_{de}
\end{equation}
with a simple power-law ansatz for $\beta(a)$, namely:
\begin{equation}
\label{cas32}
\beta(a)=\beta_0 a^\xi.
\end{equation}
Substituting this interaction form into conservation equations for DM and DE:
\begin{eqnarray}\label{eom1}
\dot{\rho}_{dm}+3H\rho_{dm}=Q,
\end{eqnarray}
\begin{eqnarray}\label{eom2}
\dot{\rho}_{de}+3H(\rho_{de}+p_{de})=-Q,
\end{eqnarray}
we get
 \begin{equation}
\label{rhophi2}
\rho_{de}=\rho_{de0}\, a^{-3(1+w_0)}\cdot
\exp{\left[\frac{3\beta_0(1-a^\xi)}{\xi}\right]},
\end{equation}
where the integration constant $\rho_{de0}$ is value of the dark energy at present,
and the dark energy EoS parameter $w\equiv p_{de}/\rho_{de}$ is a constant-$w_0$.
Substituting Eq. (\ref{rhophi2}) into Eq. (\ref{eom2}), we get
the dark matter energy density,
 \begin{equation}\label{rhom2}
\rho_{dm}=f(a)\rho_{dm0},
\end{equation}
where
\begin{equation}
\label{f1}
f(a)\equiv \frac{1}{a^3}\left\{1-\frac{\Omega_{de0}}{\Omega_{dm0}}\frac{3\beta_0 a^{-3w_0}e^{\frac{3\beta_0}{\xi}}}{\xi}\cdot\left[a^\xi E_{\frac{3w_0}{\xi}}\left(\frac{3\beta_0 a^\xi}{\xi}\right)-a^{3w_0} E_{\frac{3w_0}{\xi}} \left(\frac{3\beta_0}{\xi}\right)\right]\right\},
\end{equation}
where $\rho_{dm0}$ is dark matter density at present day, and $E_n(z)=\int_1^\infty t^{-n}e^{-xt}dt$ the
usual exponential integral function.
Note however that Eq. (\ref{rhom2}) is an
analytical expression, while in the corresponding expressions were left as
integrals and were calculated numerically. Obviously, in the case of
non-interaction (that is, for $\beta_0=0$), Eq. (\ref{rhom2}) recovers the standard
result $\rho_{dm}=\rho_{dm0}/a^3$.
For the special case $\xi=0$ find dimensionless Hubble parameter $E^2(z)\equiv \frac{H^{2}}{H^{2}_0}$, the evolution of the density parameters $\Omega_b(a)$, $\Omega_{dm}(a)$ and $\Omega_{de}(a)$ and $q(a)$. For what values of $\beta_0$ the cosmic acceleration is transient?

\item Consider the flat FLRW cosmology with two coupled
homogeneous scalar fields $\Phi$ and $\Psi$:
\begin{eqnarray}
\dot{\rho_b} &=& -3 H \gamma_b \rho_b \\
\ddot{\Phi} &=& -3H\dot{\Phi} - \partial_\Phi V \\
\ddot{\Psi} &=& -3H\dot{\Psi} - \partial_\Psi V \\
\dot{H} &=& -4\pi G(\gamma_m\rho_m+\gamma_r\rho_r+\gamma_Q\rho_Q)~,
\end{eqnarray}
\begin{equation}
H^2 = \frac{8\pi G}{3}(\rho_m+\rho_r+\rho_Q)-\frac{k}{a^2}~.
\end{equation}
Here a dot denotes a derivative with respect to the cosmic time $t$, the subscript $b$
refers to the dominant background quantity, either dust (m) or radiation (r) while $Q$
refers to the Dark Energy sector, here the two quintessence scalar fields.

The quintessence fields with potential $V$ have the following
energy density and pressure:

\begin{eqnarray}
  \rho_Q =\frac{1}{2}\dot{\Phi}^2 + \frac{1}{2}\dot{\Psi}^2 + V(\Phi,\Psi)\\
  p_Q =\frac{1}{2}\dot{\Phi}^2 + \frac{1}{2}\dot{\Psi}^2 - V(\Phi,\Psi)
\end{eqnarray}
with $p_Q = (\gamma_Q-1)\rho_Q$.
It is convenient to define the following new variables :
\begin{equation}
X_{\Phi}=\sqrt{\frac{8\pi G}{3H^2}}~\frac{\dot{\Phi}}{\sqrt{2}},~~~
X_{\Psi}=\sqrt{\frac{8\pi G}{3H^2}}~\frac{\dot{\Psi}}{\sqrt{2}},~~~
X_V=\sqrt{\frac{8\pi G}{3H^2}}~\sqrt{V}.
\end{equation}
Find expressions for the $\Phi'$, $\Psi'$, $X_{\Phi}'$, $X_{\Psi}'$, $X_V'$ where a prime denotes a derivative with respect to the quantity $N$, the number of e-folds with respect to the present time,
\begin{equation}
N\equiv {\rm ln} \frac{a}{a_0}~,\label{N}
\end{equation}
and we have also $H=\dot{N}$.

  \item Using result from the previous problem find the relative energy density for matter, radiation and quintessence,  $\Omega_m$, $\Omega_r$ and $\Omega_Q$, the deceleration parameter $q$, the Hubble-parameter-free luminosity distance $D_L$ and the age of the Universe $t_0$.

\end{enumerate}

\end{document}